\documentstyle[preprint,aps,psfig,epsfig,floats,amssymb]{revtex}
\tightenlines
\def\:{\hbox{\bf :}}\def\d{\mbox{d}}
\begin{document}
\title{Universal homodyne tomography with a single local oscillator}
\author{G. Mauro D'Ariano and  Massimiliano F. Sacchi}
\address{Dipartimento di Fisica ``A. Volta'', Universit\`a di Pavia
and INFM,\\ via A. Bassi 6, I-27100 Italy}
\author{Prem Kumar}
\address{Department of Electrical and Computer Engineering,
Northwestern University,\\ Evanston, IL 60208, USA} 
\date{\today}
\maketitle
\begin{abstract}
We propose a general method for measuring an arbitrary observable of a
multimode electromagnetic field using homodyne detection with a
single local oscillator. In this method the 
local oscillator scans over all 
possible linear combinations of the modes. The case of two modes is 
analyzed in detail and the feasibility of the measurement is studied
on the basis of Monte-Carlo simulations. We also provide an
application of this method in tomographic testing of the GHZ state. 
\end{abstract}
\section{Introduction}
Optical homodyne tomography is a well-established quantitative method
for measuring the quantum state of radiation and for obtaining the
expectation value of arbitrary observables of the field
\cite{raymer,raymer95,breiten} (for a review see Ref.~\cite{bilk}).
The success of optical homodyne tomography has stimulated research
relating to the state-reconstruction procedures in other fields, such
as in the realm of atomic \cite{12}, molecular \cite{13}, and ion-trap
\cite{15} physics.  As a matter of fact, the tomographic method is a
kind of universal detection technique \cite{univ}, with which one can measure
any observable $ O$ of the field by averaging a suitable unbiased
estimator ${\cal E} [ O](x,\phi )$ over the homodyne data $x$ at
random phase values $\phi $. Single-mode homodyne tomography can be
immediately generalized to multimode fields. For factorized multimode
operators $ O= O_1\otimes O_2\otimes \ldots  \otimes O_n$ the
corresponding estimator is just the product of the estimators for each
of the single-mode operators $ O_1,O_1,\ldots, O_n$. By linearity the
estimator can then be extended to generic multimode
operators. However, such a simple generalization requires a separate
homodyne measurement for each of the modes, which cannot be achieved
in practice when the 
modes of the field are not spatio-temporally separated. For this reason,
tomographic methods have been devised which either use only a single
local oscillator (LO) \cite{Leonhardt}, or avoid the use of conventional
homodyne detection \cite {our}.  However, both the methods work for
only two modes of the field, and the self-homodyne method of
Ref.~\cite{our} is suitable only in special experimental situations
(e.g., in the tomography of parametrically downconverted
radiation). Therefore, a more general multimode tomographic method is
needed, especially in consideration of the possibility of a precise
analysis for pulsed fields, for which the problem of mode matching
between the LO and the detected fields (determined
by their relative spatio-temporal overlap) \cite{matched-lo} gives a
detrimental contribution to the overall quantum efficiency.  \par In
this paper we propose a general method for measuring an arbitrary
observable of the multimode electromagnetic field, which uses homodyne
detection with a {\em single} LO.  We provide the rule for evaluating
the ``unbiased estimator'' of a generic multimode operator. The
quantum expectation value of the operator can then be obtained for any
unknown state of the radiation field through an average of this
estimator over the homodyne outcomes that are collected using a single
LO which scans over different linear combinations of the incident
modes. The paper is organized as follows: In Sec.~II we present the
general method for obtaining the estimator pertaining to an arbitrary
multimode operator. Upon averaging this estimator over the homodyne
outcomes, one obtains the quantum expectation value of the
corresponding operator. We specialize to observables corresponding to
the matrix elements of the multimode density operator and to the total
number of photons. In the two-mode case we explicitly derive the
estimator for the four-dimensional Q-function and for the moments
generating function of the total number of photons. In Sec.~III we
investigate the experimental conditions for extracting the joint
photon-number probability and the distribution of the total number of
photons for two-mode quantum states. We present the results of some
Monte-Carlo simulations for the twin-beam state that is produced by
nondegenerate parametric amplification (spontaneous
downconversion). We average the estimators obtained in Sec.~II over
the homodyne data that are distributed according to the theoretical
homodyne probability evaluated in the Appendix. The simulations show
that the measurement is feasible for quantum efficiency values of the
homodyne detector in the $80$--$90\%$ range and with the number of
experimental data samples of order $10^6$--$10^7$. In Sec.~IV we show an
application of our method in measurement of the three-particle
maximally-entangled state called the GHZ state. In such a case the
number of radiation modes is six and a more suitable arrangement of
the tomographic machine requires the use of three LO's. The results of
Monte-Carlo simulations show that for homodyne detectors with
quantum efficiency value $\eta =85\%$ one needs about $10^7$ data
samples to reconstruct the state with a relatively small statistical
error.  Finally, some conclusions are drawn in Sec.~V.
\section{The general method}
For a single-mode radiation field one has the following resolution of
the identity on the Hilbert-Schmidt space: 
\begin{eqnarray}
  O=\int
\frac{\d^2z}{\pi}\mbox{Tr}[  O D^{\dag}(z)]
D(z)\;,
\label{tomoid}
\end{eqnarray}
where $ O$ is a Hilbert-Schmidt operator, the integral is extended to
the complex plane ${\mathbb C}$ for $z$, and $
D(z)=\exp(za^{\dag}-z^*a)$ denotes the displacement operator for the
field mode with annihilation and creation operators $a$ and
$a^{\dag}$, respectively, having the commutation relation
$[a,a^{\dag}]=1$. Equation~(\ref{tomoid}) simply follows from the
orthogonality relation for displacement operators $\hbox{Tr}[ D(z)
D^{\dagger}(z')]=\delta_{2}(z-z')$, $\delta_{2}(z)$ denoting the Dirac
delta-function on the complex plane.  Equation (\ref{tomoid}) is the
starting point of our method; it can be easily generalized to any
number of modes as follows:
\begin{eqnarray}
O&=&\int\frac{\d^2z_0}{\pi}\int\frac{\d^2z_1}{\pi}\ldots\int\frac{\d^2z_M}{\pi}
\mbox{Tr}\left\{  O \exp\left[\sum_{l=0}^M\left(-z_l
a_l^{\dag}+z_l^* a_l\right)\right]\right\} \nonumber \\&\times & 
\exp\left[\sum_{l=0}^M\left(
z_l a_l^{\dag}-z_l^* a_l\right)\right] \;,\label{tomoidmany}
\end{eqnarray}
where $a_l$ and $a_l^{\dag}$, with $l=0,\ldots,M$ and
$[a_l,a_{l'}^{\dag}]=\delta_{ll'}$, are the annihilation and creation
operators, respectively, of the $M+1$ independent modes, and $O$ now 
denotes an operator over all the modes. Using the following hyper-spherical
parameterization for $z_l\in\mathbb C$
\begin{eqnarray}
z_0=\frac i2 k\, u_0({\vec\theta})e^{i\psi_0}&\doteq& \frac i2 k\, e^{i\psi_0}\cos\theta_1\;,\nonumber\\
z_1=\frac i2 k\, u_1({\vec\theta})e^{i\psi_1}&\doteq& \frac i2 k\, e^{i\psi_1}\sin\theta_1\cos\theta_2\;,\nonumber\\
z_2=\frac i2 k\, u_2({\vec\theta})e^{i\psi_2}&\doteq&
\frac i2 k\, e^{i\psi_2}\sin\theta_1\sin\theta_2\cos\theta_3\;,\nonumber\\ 
&\ldots&\nonumber\\ 
z_{M-1}=\frac i2 k\, u_{M-1}({\vec\theta})e^{i\psi_{M-1}}&\doteq&
\frac i2 k\, e^{i\psi_{M-1}}\sin\theta_1\sin\theta_2\ldots\sin\theta_{M-1}\cos\theta_M\;,\nonumber\\
z_M=\frac i2 k\, u_M({\vec\theta})e^{i\psi_M}&\doteq&
\frac i2 k\, 
e^{i\psi_M}\sin\theta_1\sin\theta_2\ldots\sin\theta_{M-1}\sin\theta_M\;,
\label{hyper}
\end{eqnarray}
where $k\in[0,\infty)$; $\psi_l\in[0,2\pi]$ for $l=0,1,\ldots ,M$; 
and $\theta_l\in[0,\pi/2]$ for $l=1,2,\ldots,M$, Eq.~(\ref{tomoidmany}) can be
rewritten as follows: 
\begin{eqnarray} 
O=\int\d\mu[{\vec\psi}]\int\d\mu[{\vec\theta}]\int_0^{+\infty}\d k\,
\left(\frac{k}{2}\right)^{2M+1}\frac{1}{M!}
\mbox{Tr}\{  O\exp[-ik  X({\vec\theta},{\vec\psi})]\}
\exp[ik  X({\vec\theta},{\vec\psi})]\;.\label{thpsi1}
\end{eqnarray}
Here we have used the notation
\begin{eqnarray}
\int\d\mu[{\vec\psi}]\doteq\prod_{l=0}^M\int_0^{2\pi}\frac{\d\psi_l}
{2\pi}\;,\qquad
\int\d\mu[{\vec\theta}] \doteq 2^M
\,M!\prod_{l=1}^M\int_0^{\pi/2}d\theta_l\,
\sin^{2(M-l)+1}\theta_l\cos\theta_l\;,
\label{notaz}
\end{eqnarray}
\begin{eqnarray}
  X({\vec\theta},{\vec\psi})&=&{1\over2}\left[ 
A^{\dag}({\vec\theta},{\vec\psi})+ 
A({\vec\theta},{\vec\psi})\right]\;,\label{X}\\  
A({\vec\theta},{\vec\psi})&=& \sum_{l=0}^M
e^{-i\psi_l}u_l({\vec\theta})a_l\;.
\end{eqnarray}
Notice that, thanks to the parameterization in Eq.~(\ref{hyper}),
where $\sum_{l=0}^{+\infty}u^2_l({\vec\theta})=1$, one has the
commutation relation $[ A({\vec\theta},{\vec\psi}),
A^{\dag}({\vec\theta},{\vec\psi})]=1$, which implies that
$A({\vec\theta},{\vec\psi})$ and $ A^{\dag}({\vec\theta},{\vec\psi})$
themselves are annihilation and creation operators, respectively, of a
bosonic mode. Also, by scanning all values of $\theta_l\in[0,\pi/2]$
and $\psi_l\in[0,2\pi]$, all possible linear combinations of the modes
described by annihilation operators $a_l$, with $l=0,\ldots,M$, are
obtained.  \par For a single mode of the radiation field the
experimental homodyne probability distribution of a field quadrature
with quantum efficiency $\eta<1$ is a Gaussian convolution with
variance $\Delta^2_{\eta}=\frac{1-\eta}{4\eta}$ of the ideal
probability distribution. Therefore, for the quadrature operator $
X({\vec\theta},{\vec\psi})$ in Eq.~(\ref{X}), one has the following
identity for the moments generating function
\begin{eqnarray}
\langle e^{ik  X}\rangle=\exp\left(\frac{1-\eta}{8\eta}k^2\right)
\int_{-\infty}^{+\infty}\d x\, e^{ikx}\,p_{\eta}(x;{\vec\theta},{\vec\psi})
\;,\label{momM}
\end{eqnarray}
where $p_{\eta}(x;{\vec\theta},{\vec\psi})$ denotes the homodyne
probability distribution of the quadrature $
X({\vec\theta},{\vec\psi})$ with quantum efficiency $\eta$.
Generally, $\eta $ can depend on the mode itself, i.e., it is a
function $\eta=\eta({\vec\theta},{\vec\psi})$ of the selected mode. In
the following, for simplicity, we assume $\eta$ to be mode
independent, however. By taking the ensemble average on each side of
Eq.~(\ref{thpsi1}) and using Eq.~(\ref{momM}) one has
\begin{eqnarray}
\langle  O\rangle=\int\d\mu[{\vec\psi}]\int\d\mu[{\vec\theta}]\,
\int _{-\infty}^{+\infty}\d x \,p_\eta (x;{\vec\theta},{\vec\psi})
\,{\cal E}_{\eta}[  O](x;{\vec\theta},{\vec\psi})\;,\label{K}
\end{eqnarray}
where, for a given operator $  O$, the function ${\cal E}
_{\eta}[  O](x;{\vec\theta},{\vec\psi})$ of $x$, $\vec\theta $,
$\vec\psi $ has the following analytic expression
\begin{eqnarray}
{\cal E}_{\eta}[  O](x;{\vec\theta},{\vec\psi})=\frac{\kappa^{M+1}}{M!}
\int_0^{+\infty} \d t\, e^{-t+2i\sqrt{\kappa t}\,x}\, t^M\, \mbox{Tr}\{ 
O\,\:\exp[-2i\sqrt{\kappa t}  X({\vec\theta},{\vec\psi})]\:\}
\;\label{MAIN}
\end{eqnarray}
with
\begin{eqnarray}
\kappa=\frac{2\eta}{2\eta-1}\;.\label{kappa}
\end{eqnarray}
Equation~(\ref{MAIN}) is the central result of this paper. For any
given operator $  O$ it provides the ``unbiased estimator'' to be
averaged over all homodyne outcomes of the quadrature $ 
X({\vec\theta},{\vec\psi})$ in order to obtain the ensemble average
$\langle  O\rangle$ for any unknown state of the radiation field.  The
homodyne outcomes for $  X({\vec\theta},{\vec\psi})$ can be obtained
by using a single LO that is prepared in the multimode coherent
state $\otimes _{l=0}^M |\gamma _l \rangle $ with 
$\gamma _l =e^{i\psi _l}u_l(\theta) K/2$ and  $K \gg 1$. 
In fact, in this case the rescaled zero-frequency photocurrent at the 
output of a balanced homodyne detector is given by 
\begin{eqnarray}
  i =\sum_{l=0}^M(\gamma ^*_l a_l+\gamma _l a_l^\dag )/K
\;,\label{dcc}
\end{eqnarray}
which corresponds to the operator $ X({\vec\theta},{\vec\psi})$. In
the limit of a strong LO ($K\rightarrow \infty$), all
moments of the current $ i$ correspond to the moments of $
X({\vec\theta},{\vec\psi})$, and the exact measurement of $
X({\vec\theta},{\vec\psi})$ is then realized.  
Notice that for modes $a_l$ with different frequencies, in the d.c. 
photocurrent in Eq. (\ref{dcc}) each LO with amplitude $\gamma _l$ 
selects the mode $a_l$ at the same frequency (and polarization). 
For the effect of
less-than-unity quantum efficiency, previous considerations on
Eq.~(\ref{momM}) apply.  \par In order to obtain the ensemble average
in Eq.~(\ref{K}) as an experimental average over the homodyne
outcomes, one needs to satisfy the validity conditions of the
central-limit theorem. Since in the strong LO 
approximation the probability $p_{\eta}(x;{\vec\theta},{\vec\psi})$
must decay as a Gaussian for large $x$, it follows that the integral
in Eq.~(\ref{K}) can be experimentally sampled for any {\em a priori}
unknown probability distribution $p_{\eta}(x;{\vec\theta},{\vec\psi})$
if ${\cal E}_{\eta}[ O](x;{\vec\theta},{\vec\psi})$ increases slower
than $\exp(\kappa x^2)$ for large $x$, and is bounded for
$|x|<+\infty$. In this case one is guaranteed that the integral in
Eq.~(\ref{K}) can be statistically sampled over a sufficiently large
set of data. The average values for different experiments will be
Gaussian distributed around the mathematical expectation in
Eq.~(\ref{K}), allowing estimation of the confidence intervals, which
will decrease as the inverse square-root of the number of experimental
data. In general, the boundedness of ${\cal E}_{\eta}[
O](x;{\vec\theta},{\vec\psi})$ for $|x|<+\infty$ will pose lower
bounds for the quantum efficiency $\eta$ below which the measurement
cannot be performed, similarly to what happens in the one-mode case
\cite{ulf}. This limitation is due to the generality of the method,
which is perfectly unbiased, and makes no {\em a priori} assumption on
the state of the radiation field, the only approximation being that of
a strong LO. This should be contrasted with other
methods, as the maximum entropy method \cite{buzek} or the maximum
likelihood method \cite{banaszek,hradil}, which do not suffer such
limitation on the quantum efficiency; however they are generally
biased and based on assumptions for the state of the radiation field.  \par
Equation~(\ref{K}) can be specialized to some observables $ O$ of
interest. In particular, one can estimate the matrix element
$\langle\{n_l\}| R|\{m_l\}\rangle$ of the multimode density operator 
$R$. This will be obtained by averaging the following estimator:
\begin{eqnarray}
{\cal E}_{\eta}[|\{m_l\}\rangle\langle\{n_l\}|](x;{\vec\theta},{\vec\psi})&=&
e^{-i\sum_{l=0}^M(n_l-m_l)\psi_l}\,
\frac{\kappa^{M+1}}{M!}
\prod_{l=0}^M\left\{[-i\sqrt{\kappa} u_l({\vec\theta})]^{\mu_l-\nu_l}
\sqrt{\frac{\nu_l!}{\mu_l!}}\right\}\nonumber\\ &\times&
\int_0^{+\infty}\d t\,e^{-t+2i\sqrt{\kappa t}\,x}\,
t^{M+\sum_{l=0}^M(\mu_l-\nu_l)/2}\prod_{l=0}^ML_{\nu_l}^{\mu_l-\nu_l}
[\kappa u_l^2({\vec\theta})t]\;,\label{gasp}
\end{eqnarray}
where $\mu_l=\mbox{max}(m_l,n_l)$, $\nu_l=\mbox{min}(m_l,n_l)$, and
$L_n^{\alpha}(z)$ denotes the customary generalized Laguerre
polynomial of variable $z$. For diagonal matrix elements,
Eq.~(\ref{gasp}) simplifies to 
\begin{eqnarray}
{\cal E}_{\eta}[|\{n_l\}\rangle\langle\{n_l\}|](x;{\vec\theta},{\vec\psi})&=&
\frac{\kappa^{M+1}}{M!}
\int_0^{+\infty}\d t\,e^{-t+2i\sqrt{\kappa t}\,x}\,
t^M\prod_{l=0}^ML_{n_l}[\kappa u_l^2({\vec\theta})t]\;\label{minigasp}
\end{eqnarray}
with $L_n(z)$ denoting the customary Laguerre polynomial in $z$. 
Notice that the estimator in Eq.~(\ref{minigasp}) does not depend on
the phases $\psi_l$; only the knowledge of the angles $\theta_l$ is needed.
Using the following identity for the Laguerre polynomials \cite{gradstein}:
\begin{eqnarray}
L_n^{\alpha_0+\alpha_1+\ldots+\alpha_M+M}(x_0+x_1+\ldots+x_M)=
\sum_{i_0+i_1+\ldots+i_M=n}
L_{i_0}^{\alpha_0}(x_0)L_{i_1}^{\alpha_1}(x_1)\ldots
L_{i_M}^{\alpha_M}(x_M)\;,
\end{eqnarray}
from Eq.~(\ref{minigasp}) one can easily derive the estimator for the
probability distribution of the total number of photons $ 
N=\sum_{l=0}^Ma^{\dag}_la_l$ 
\begin{eqnarray}
{\cal E}_{\eta}[|n\rangle\langle n|](x;{\vec\theta},{\vec\psi})&=&
\frac{\kappa^{M+1}}{M!}
\int_0^{+\infty}\d t\,e^{-t+2i\sqrt{\kappa t}\,x}\,
t^M L^{M}_n[\kappa t]\;,\label{N}
\end{eqnarray}
where $|n\rangle$ denotes the eigenvector of $ N$ with eigenvalue $n$.
Notice that the estimator in Eq.~(\ref{N}) does not depend on any of
the phases $\psi_l$ or the angles $\theta_l$, and thus their knowledge
is not needed in the measurement of the probability distribution of
the total number of photons. \par Now we specialize to the case of
only two modes (i.e., M=1 and $\vec \theta$ is a scalar $\theta $).  
The joint photon-number probability distribution is
obtained by averaging the following estimator:
\begin{eqnarray}
{\cal E}_{\eta}[|n,m\rangle\langle n,m|](x;\theta,\psi_0,\psi_1)=
\kappa^2\int_0^{+\infty}\d t\,e^{-t+2i\sqrt{\kappa t}\,x}\,
t \,L_n(\kappa t\cos^2\theta)L_m(\kappa t\sin^2\theta)\;.\label{twogasp}
\end{eqnarray}
Analogously, using Eq.~(\ref{MAIN}) one can derive the following estimator for
the four dimensional Q-function:
\begin{eqnarray}
&&{\cal E}_
{\eta}[|\alpha,\beta\rangle\langle\alpha,\beta|](x;\theta,\psi_0,\psi_1)=
\nonumber
\\&&
\kappa^2 \Phi\left(2,{1\over2};-\kappa\left[x-{1\over2}\cos\theta\,\mbox{Im}
(\alpha^*e^{i\psi_0})
-{1\over2}\sin\theta\,\mbox{Im}(\beta^*e^{i\psi_1})\right]^2\right)
\;,\label{Q}
\end{eqnarray}
where $|\alpha,\beta\rangle$ with $\alpha,\beta\in\mathbb C$ denotes a
two-mode coherent state, and $\Phi(a,b;z)$ is the customary confluent
hypergeometric function of $z$. The estimator
(\ref{N}) for the probability distribution of the total number of
photons can be written as
\begin{eqnarray}
{\cal E}_{\eta}[|n\rangle\langle n|](x;\theta,\psi_0,\psi_1)=
\kappa^2\int_0^{+\infty}\d t\,e^{-t+2i\sqrt{\kappa t}\,x}\,
t\,L^1_n[\kappa t]\;.\label{N1}
\end{eqnarray}
For the total number of photons one can also derive the
estimator for the moment generating function, using the generating
function for the Laguerre polynomials \cite{gradstein}. One obtains
\begin{eqnarray}
{\cal E}_{\eta}[z^{a^{\dag}a+b^{\dag}b}](x;\theta,\psi_0,\psi_1)= 
\frac{1}{(z+\frac{1-z}{\kappa})^2}\Phi\left(2,{1\over2};-
\frac{1-z}{z+\frac{1-z}{\kappa}}\,x^2\right)\;,\label{gen}
\end{eqnarray}
where we have denoted by $a$ and $b$ the annihilation operators of the two
modes. For the first two moments one obtains the simple expressions
\begin{eqnarray}
{\cal E}_{\eta}[a^{\dag}a+b^{\dag}b](x;\theta,\psi_0,\psi_1)&=& 
4x^2+{2\over{\kappa}}-2\;,\\
{\cal E}_{\eta}[(a^{\dag}a+b^{\dag}b)^2](x;\theta,\psi_0,\psi_1)&=& 
8x^4+\left({24\over{\gamma}}-20
\right)x^2+{6\over{\gamma^2}}-{10\over{\gamma}}+4\;.
\end{eqnarray}
It is worth noting that analogous estimators for the photon-number 
difference between the two modes are singular and one needs a cutoff procedure,
similar to the one used in Ref.~\cite{our} for recovering the
correlation between the modes by means of the customary two-mode tomography. 
The singular behavior of the  estimators for the photon-difference
operators can be understood simply from the fact that to extract 
information pertaining to a single mode only one needs a
delta-function at $\theta=0$ for mode $a$, or $\theta=\pi/2$ for mode
$b$, and, in this case, one could better use the original one-mode
tomography method \cite{ulf} by setting the LO to the
proper mode of interest.
\par Finally, we note that for the case of two-mode tomography 
the estimator ${\cal E}_{\eta }$ can be averaged by taking the
integral
\begin{eqnarray}
\langle  O\rangle=\int_0^{2\pi}\frac{\d\psi_0}{2\pi}
\int_0^{2\pi}\frac{\d\psi_1}{2\pi}\int_{-1}^1\frac{\d(\cos 2\theta)}{2}\, 
\int_{-\infty }^{+\infty }\d x\, p_{\eta }(x;\theta,\psi_0,\psi_1)\,
{\cal E}_{\eta}[  O](x;\theta,\psi_0,\psi_1)\;\label{K1}
\end{eqnarray}
over the random parameters $\cos(2\theta),\psi_0$, and $\psi_1$. For
example, in the case of two radiation modes having the same frequency
but orthogonal polarizations, $\theta $ represents a random rotation
of the polarizations, whereas $\psi_0$ and $\psi_1$ denote the
relative phases between the LO and the two modes, respectively.
\section{Numerical results for two modes}
In this section we present some Monte-Carlo simulations in order to
estimate the working experimental conditions for performing the 
single-LO tomography on two-mode fields. We focus our attention
on the twin-beam state, usually generated by spontaneous parametric 
downconversion, namely 
\begin{eqnarray}
|\Psi\rangle=(1-|\xi|^2)^{{1\over2}}\sum_{n=0}^{\infty}\xi^n\,
|n\rangle_a|n\rangle_b\;.\label{Psi}
\end{eqnarray}
For the simulations we use the following homodyne probability
distribution that is derived in the Appendix:
\begin{eqnarray}
p_\eta (x;\theta,\psi_0,\psi_1)={1\over{\sqrt{2\pi\Delta^2_\eta
(\theta,\psi_0,\psi_1)}}}
\exp\left(-\frac{x^2}{2\Delta_\eta^2(\theta,\psi_0,\psi_1)}\right)
\;,\label{prob}
\end{eqnarray}
where the variance  $\Delta_\eta ^2(\theta,\psi_0,\psi_1)$ is given by
\begin{eqnarray}
\Delta_\eta ^2(\theta,\psi_0,\psi_1)=\frac{1+|\xi|^2+2|\xi|
\sin 2\theta\cos(\psi_0+\psi_1-\arg\xi)}
{4(1-|\xi|^2)}+\frac{1-\eta}{4\eta}\;.\label{delta}
\end{eqnarray}
In the case of two radiation modes having the same frequency but
orthogonal polarizations, Eq.~(\ref{prob}) gives the theoretical
probability of outcome $x$ for the homodyne measurement at a 
polarization angle $\theta $, $\psi_0$ and $\psi_1$ denoting the
relative phases between the LO and the two modes, respectively.
\par We study the tomographic measurement of the joint photon-number
probability distribution and the probability distribution for the
total number of photons with use of the estimators in Eqs.~(\ref{twogasp})
and (\ref{N1}), respectively. Moreover, we reconstruct the matrix elements 
\begin{eqnarray}
C_{n,m}\equiv {}_a\langle m|{}_b \langle m|\Psi \rangle \langle 
\Psi |n \rangle _a |n \rangle _b  
\;\label{cnm}
\end{eqnarray}
that reveal the coherence of the twin-beam state by using the
estimator in Eq.~(\ref{gasp}). For the twin-beam state in
Eq.~(\ref{Psi}), one should have 
\begin{eqnarray}
C_{n,m}=(1-|\xi |^2)\xi^m\,\xi^{*n}
\;.\label{cnmxi}
\end{eqnarray} 
The estimators have been numerically
evaluated by applying the Gauss method for calculating the integral in
Eq.~(\ref{gasp}), which results in a fast and sufficiently precise
algorithm with use of just 150 evaluation points. Notice that in the
present case there is no convenience in using the factorization
formula given in Ref.~\cite{Leonhardt}, as in that case an integral of
a product of functions is needed. \par
\begin{figure}[hbt]\begin{center}
\epsfxsize=.47\hsize\leavevmode\epsffile{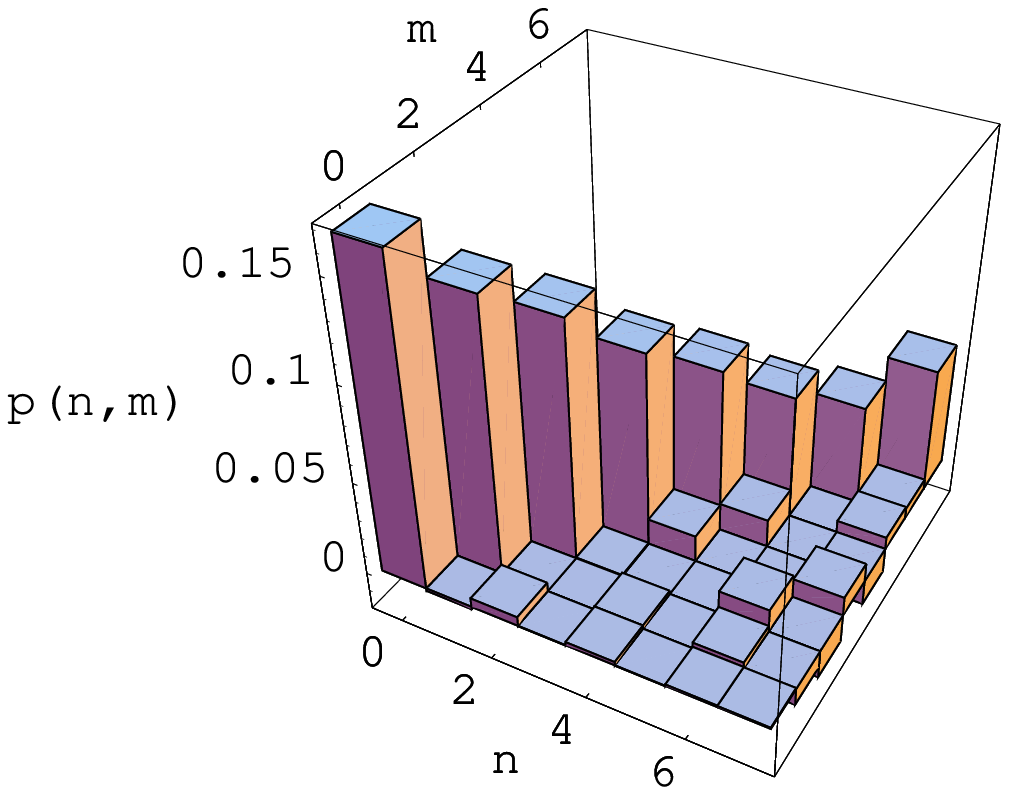}\hspace{12pt}
\epsfxsize=.47\hsize\leavevmode\epsffile{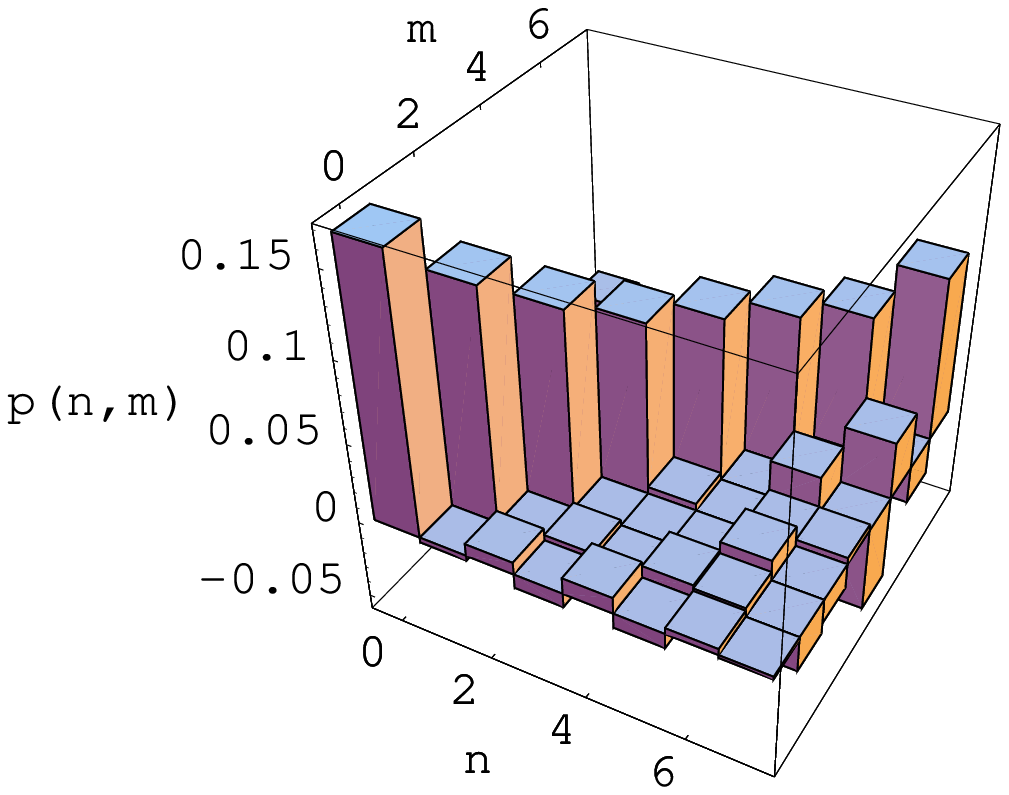}\end{center}
\vskip 1truecm
\caption{Two-mode photon-number probability $p(n,m)$ of the twin-beam state
of parametric fluorescence in Eq.~(\ref{Psi}) for average number of
photons per beam $\overline{n}=|\xi|^2/(1-|\xi|^2)=5$ obtained by a
Monte-Carlo simulation of the probability in Eqs. (\ref{prob}) and
(\ref{delta}) with random parameters $\cos 2\theta$, $\psi_0$, and
$\psi_1$, using the estimator in Eq.~(\ref{twogasp}). On the left we
have quantum efficiency $\eta=1$ and $10^6$ data samples were used in
the reconstruction. 
On the right quantum efficiency $\eta=0.9$, and 
$5\times 10^6$ data samples were used. The theoretical values of off-diagonal 
$p(n,m)$ are zero; for a comparison between theoretical and
experimental diagonal $p(n,n)$ probabilities and their relative statistical
errors, see analogous experiments in Figs. \ref{f:sum1} and \ref{f:sum2}.} 
\label{f:matrix}\end{figure}

 In Fig. \ref{f:matrix} a
Monte-Carlo simulation of the joint photon-number probability
distribution is presented. The simulated values compare very well with
the theoretical ones. We have done a careful analysis of the
statistical errors for various twin-beam states by constructing histograms of
deviations of the results from different simulated experiments 
from the theoretical ones. In comparison to
the customary two-LO tomography \cite{our}, where for $\eta=1$ the
statistical errors saturate for increasingly large $n$ and $m$, here
we have statistical errors that are slowly increasing versus $n$ and
$m$. This is due to the fact that the range of the estimator in
Eq.~(\ref{twogasp}) increases versus $n$ and
$m$. Overall we find that for any given quantum efficiency the
statistical errors are generally slightly larger than those obtained
with the two-LO method. The convenience of using a single LO then
comes with its own price tag.  \par
\begin{figure}[hbt]\begin{center}
\epsfxsize=.47\hsize\leavevmode\epsffile{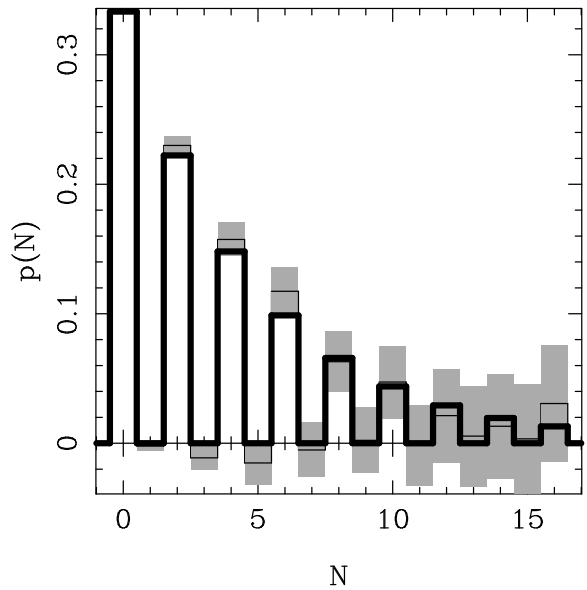}
\end{center}
\vskip 1truecm
\caption{Probability distribution for the total number of photons of
the twin-beam state in Eq.~(\ref{Psi}) for average number of
photons per beam $\overline{n}=2$ obtained 
using the estimator in Eq.~(\ref{N1}). 
The oscillation of the total photon-number probability due to the
perfect correlation of the twin-beam state has been reconstructed by 
simulating $10^6$ data samples with quantum efficiency $\eta=1$. The
theoretical probability (thick solid line) is superimposed onto the
result of the Monte-Carlo experiment; the latter is shown by the thin
solid line with the statistical errors in gray shade.}
\label{f:sum1}\end{figure}

\begin{figure}[hbt]
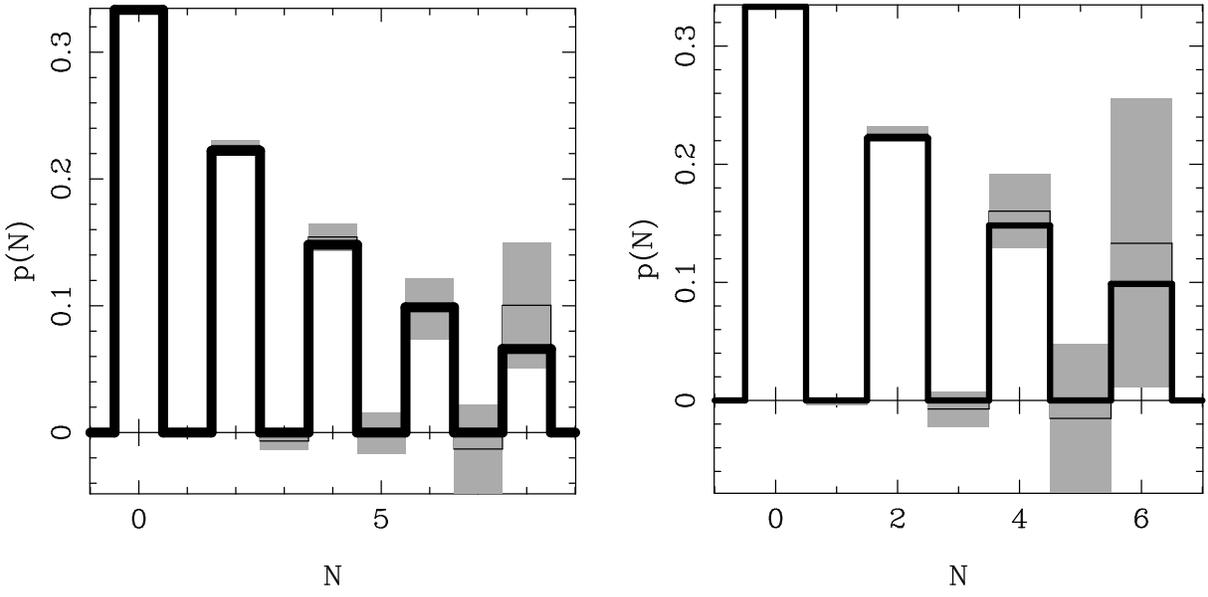
\begin{center}
\epsfxsize=.47\hsize\leavevmode\epsffile{testssum2.ps}\hspace {12pt}
\epsfxsize=.47\hsize\leavevmode\epsffile{testssum3.ps}
\end{center}
\vskip 1truecm
\caption{Similar to Fig. \protect\ref{f:sum1}, but for quantum efficiency 
$\eta=0.9$ and $10^7$ data samples (on the left), 
and $\eta=0.8$ and $2\times 10^7$ data samples (on the right). Notice the
dramatic increase of error bars (in gray shade) 
versus N and for smaller $\eta$.}
\label{f:sum2}  
\end{figure}

\begin{figure}[hbt]\begin{center}
\epsfxsize=.47\hsize\leavevmode\epsffile{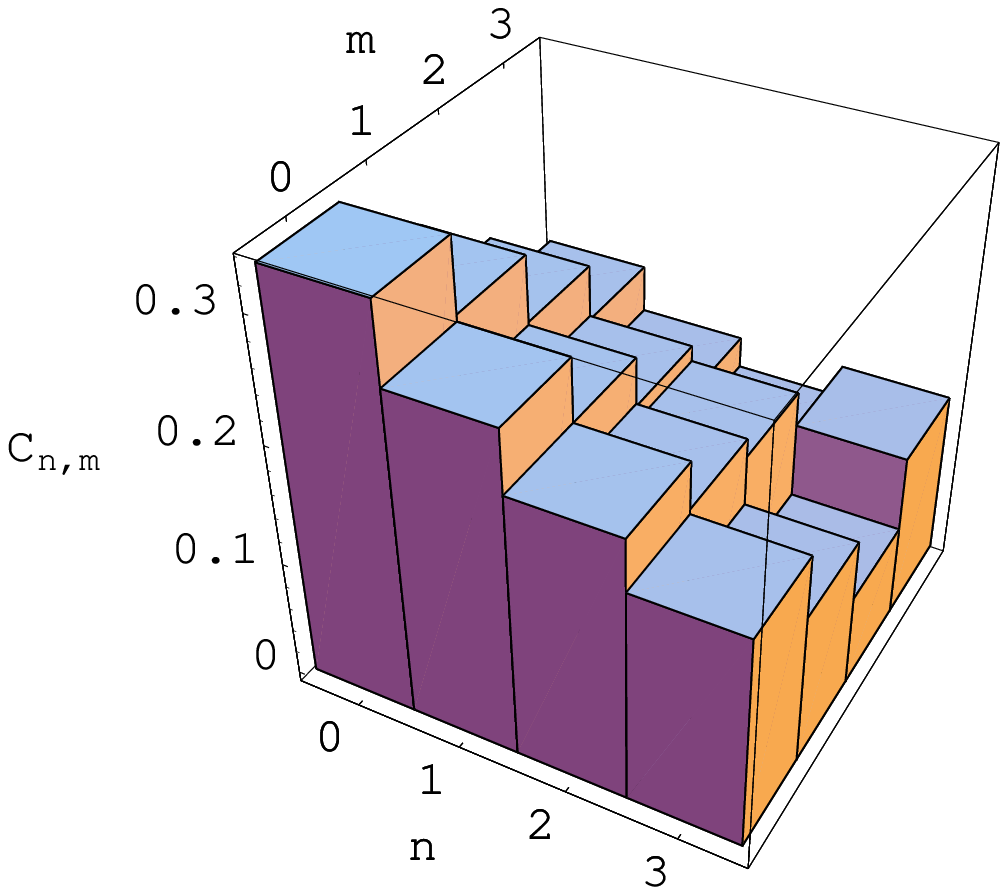}
\hspace{12pt}
\epsfxsize=.47\hsize\leavevmode\epsffile{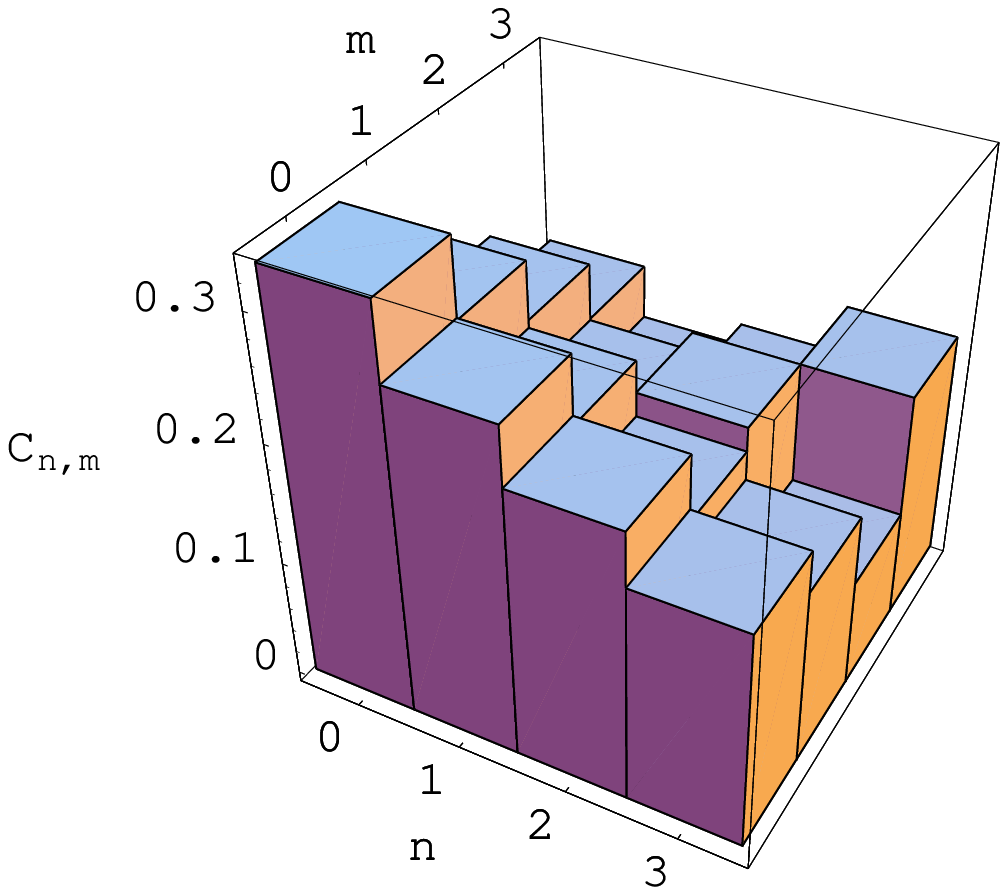}
\end{center}
\vskip 1truecm
\caption{Tomographic reconstruction of the matrix elements
$C_{n,m}\equiv {}_a\langle m|{}_b \langle m|\Psi \rangle \langle \Psi
|n \rangle _a |n \rangle _b $ of the twin-beam state of parametric
fluorescence in Eq.~(\ref{Psi}) for average number of photons per beam
$\overline{n}=2$, obtained using the estimator in Eq.~(\ref{gasp}). On
the left we used $10^6$ simulated data samples and quantum efficiency
$\eta=0.9$; on the right $3\times 10^6$ data samples and $\eta=0.8$. The
coherence of the twin-beam state is easily recognized as $C_{n,m}$
varies little for $n+m=\mbox{constant}$ [$\xi $ in Eq.~(\ref{Psi}) has
been chosen real]. For a typical comparison between 
theoretical and experimental matrix elements 
and their relative statistical
errors, see experiments in Figs. \ref{f:sum1} and \ref{f:sum2}.}
\label{f:matrix2}\end{figure}

 By using the estimator in
Eq.~(\ref{N1}) we have also constructed the probability distribution for the
total number of photons $N$ of the twin-beam state with unity
(Fig. \ref{f:sum1}) as well as less-than-unity (Fig. \ref{f:sum2}) quantum
efficiencies. Notice the dramatic increase of error bars versus N and
for smaller $\eta$. Finally, in Fig. \ref{f:matrix2} we report the
results of the tomographic measurement of the matrix elements
$C_{n,m}$ defined in Eq.~(\ref{cnm}). Because the reconstructed
$C_{n,m}$ is close to the theoretically expected value in
Eq.~(\ref{cnmxi}), these reveal the purity of the
twin-beam state, which cannot be inferred from the thermal diagonal
distribution of Fig. \ref{f:matrix}.
\section{An application to the GHZ state}
Multimode homodyne tomography allows one to verify the generation of
multimode states that are of interest in studies of the foundations of quantum 
mechanics. An example is the Greenberger-Horne-Zeilinger (GHZ) state 
\cite{citghz}, which is a $6$-mode state given by
\begin{eqnarray}
|\mbox{GHZ}\rangle \equiv \frac{1}{\sqrt 2}
\left(|1a_o\,1b_o\,1c_o \rangle -
|1a_e\,1b_e\,1c_e \rangle \right)
\;,\label{ghz}
\end{eqnarray}
wherein $o,e$ denote a couple of orthogonal polarizations; $a,b,c$
pertain to electromagnetic modes with different wavevectors and/or
frequencies; and the notation $|1,1,1 \rangle $ represents the tensor
product of three single-photon Fock states.  The GHZ state is very
interesting as it leads to correlations between three particles that
are in contradiction with the Einstein-Podolsky-Rosen idea of
``elements of reality'' \cite{epr}. We note here that no experiment
has yet succeeded in realizing the GHZ state.  \par A tomographic
measurement of the state in Eq.~(\ref{ghz}) can be suitably performed
by varying randomly the phases and polarizations of the pairs of modes
$a_{o,e}$, $b_{o,e}$, and $c_{o,e}$, and then collecting homodyne
outcomes by using three different LO's. The need of using three
separate LO's in the present case is that in the actual experimental
arrangement \cite{shih} the three beams corresponding to modes $a_{o,e}$,
$b_{o,e}$, and $c_{o,e}$ come with different wave-vectors and thus are
spatially separated.  Hence, such an experimental arrangement here
gives the opportunity of using a combination of the present multimode
method and the usual many-LO method based on the product of
single-mode estimators.  \par A simple tomographic check of the
GHZ-state production consists of measuring the expectation value on
the projector $|\phi \rangle \langle \phi |$, where
\begin{eqnarray}
|\phi \rangle \equiv \frac {1}{\sqrt 2}\left(
|1a_o\,1b_o\,1c_o \rangle +e^{i\phi }
|1a_e\,1b_e\,1c_e \rangle \right)
\;,\label{proj}
\end{eqnarray}
and comparing the result with the theoretical value, namely,
\begin{eqnarray}
C(\phi )\equiv |\langle \phi |\mbox{GHZ} \rangle |^2=\frac 12
\left(1-\cos\phi\right)\;.\label{cfi}
\end{eqnarray}
Notice that for $\phi =\pi$ the function $C(\phi)$ represents the
fidelity of the GHZ-state production. In addition, the same set of
homodyne data allows one to recover the whole interference profile in
Eq.~(\ref{cfi}) for all values of $\phi $.  \par
\begin{figure}[h]
\centerline{\psfig{file=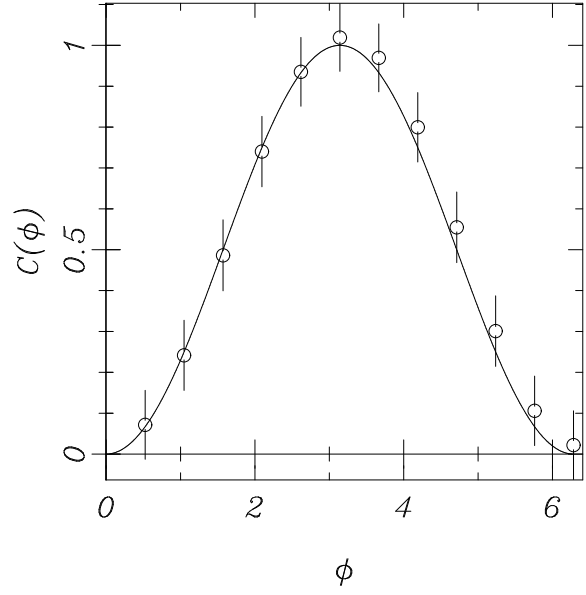,width=7.6cm}} 
\vskip 1truecm
\caption{Tomographic measurement of the overlap 
$C(\phi)$ between 
the GHZ state in Eq.~(\ref{ghz}) and  the state $|\phi \rangle $ 
in Eq.~(\ref{proj}) with varying phase $\phi $. The value for $\phi
=\pi$ represents the fidelity between the experimental state and the
theoretical one. Here a Monte-Carlo simulation with
$N=2.5\times 10^7$ data samples and quantum efficiency $\eta =0.85$. The
bars represent the statistical error, whereas the solid line is the
theoretical value of $C(\phi)$. All points are obtained from the same
data samples (which causes the evident correlation between the
statistical deviations).}\label{f:ghz}
\end{figure}

 In Fig. \ref{f:ghz}
we report the results of a Monte-Carlo simulation of the tomographic
measurement of $C(\phi )$ in Eq.~(\ref{cfi}). We used $5\times 10^7$
homodyne data samples and assumed a quantum efficiency $\eta =85\%$. For these
parameters, the simulated $C(\phi )$ compares very well with the
theoretical one.
\section{Conclusions}
We have presented a generalization of the quantum homodyne tomography
method to many modes of the radiation field that requires the use of
only a single LO.  By varying suitable random parameters the LO scans
over all the linear combinations of the field modes.  We have also
provided a general method to obtain the ``unbiased estimator'' for a
generic multimode operator. The quantum expectation value of such an
operator can be evaluated for any unknown state of the radiation field
by averaging the estimator over the homodyne outcomes that are
collected by using a single LO. The estimators for some observables,
such as the matrix elements of the multimode density operator and the
total number of photons, have been explicitly evaluated. For the
two-mode case we derived the estimator for the four-dimensional
Q-function and the moments generating function of the total number of
photons.  By means of Monte-Carlo simulations we have analyzed in
detail the case of the twin-beam state, namely the two-mode state
produced by nondegenerate parametric amplification (spontaneous
downconversion).  For quantum efficiency of homodyne detection in the
$80$--$90\%$ range and with number of data samples of order
$10^6$--$10^7$, the simulations show that measurements of the joint
photon-number probability, the distribution of the total number of
photons, and the density-matrix elements are experimentally feasible.
\par We have also shown an application of the method of multimode
homodyne tomography to the measurement of the radiation field prepared
in the Greenberger-Horne-Zeilinger state. The results of our 
simulations suggest that with a number of homodyne data samples 
around $10^7$ and a homodyne detection efficiency of $85\%$ the method
would allow the reconstruction of such an interesting state of the 
radiation field with relatively small statistical errors.
\section*{Appendix}
In this appendix we derive the theoretical probability distribution
$p_{\eta}(x;\theta,\psi_0,\psi_1)$ of the twin-beam state
\begin{eqnarray}
|\Psi\rangle=  S(\chi )|0 \rangle _a|0 \rangle_b =
(1-|\xi|^2)^{{1\over2}}\sum_{n=0}^{\infty}\xi^n|n\rangle_a|n\rangle_b\;,
\end{eqnarray}
where $  S(\chi )=\exp(\chi a^\dag b^\dag -\chi^* ab)$ and $\xi =
e^{i\arg \chi}\,\hbox{tanh}|\chi|$. 
For unity quantum efficiency, the probability density
$p(x;\theta,\psi_0,\psi_1)$ is defined as follows:
\begin{eqnarray}
p(x;\theta,\psi_0,\psi_1)&=&
\mbox{Tr}[ U^\dag \,|x\rangle_a {}_a\langle x|\otimes 1_b\,  U\,
|\Psi \rangle  \langle\Psi| ]\nonumber \\&=&
{}_a\langle 0|{}_b\langle 0|  \,S^{\dag }(\chi )
  \,U^\dag \,|x \rangle {}_a {}_a\langle x|\otimes 1_b
  \,U \,  S(\chi)\,
|0\rangle_a|0\rangle_b\label{a1}\;,
\end{eqnarray}
where $|x\rangle_a$ is the eigenvector of the quadrature $ 
x={1\over2}(a^{\dag}+a)$ with eigenvalue $x$ and 
$U$ is the unitary 
operator achieving the mode transformation
\begin{eqnarray}
  U^{\dag}{a\choose b}  U= \left(\matrix{e^{-i\psi_0}\cos\theta
&e^{-i\psi_1}\sin\theta \cr -e^{i\psi_1}\sin\theta
&e^{i\psi_0}\cos\theta \cr}\right){a\choose b}\;.\label{M}
\end{eqnarray}
In the case of two radiation modes having the same frequency but
orthogonal polarizations---the case of Type II phase-matched parametric
amplifier---Eq.~(\ref{prob}) gives the theoretical probability of
outcome $x$ for the homodyne measurement at a polarization angle $\theta
$ with respect to the polarization of the $a$ mode, and with 
$\psi_0$ and $\psi_1$ denoting  the relative phases between the LO
and the two modes, respectively. 
By using the Dirac-$\delta$ representation of the $  X$-quadrature 
projector  
\begin{eqnarray}
|x \rangle \langle x|=\int_{-\infty}^{+\infty} 
\frac{\d \lambda}{2\pi}\,\exp[i
\lambda (  X -x)] \;, 
\end{eqnarray}
Eq.~(\ref{a1}) can be rewritten as follows:
\begin{eqnarray}
&&p(x;\theta,\psi_0,\psi_1)= \int_{-\infty}^{+\infty}\frac{\d
\lambda}{2\pi}\, {}_a\langle 0|{}_b\langle 0| \,S^{\dag }(\chi)\,  
U^\dag \,e^{i \lambda (  X_a-x)}   \,U \,  S(\chi)\,
|0\rangle_a|0\rangle_b=\int_{-\infty}^{+\infty}\frac{\d
\lambda}{2\pi}\,e^{-i \lambda x}\times  \label{a2} 
\\&& {}_a\langle
0|{}_b\langle 0| \exp\left\{i \frac \lambda 2 \left[(e^{-i\psi_0}\mu
\cos\theta +e^{i\psi_1}\nu^*\sin\theta )a+ (e^{i\psi_0}\nu^*
\cos\theta +
e^{-i\psi_1}\mu\sin\theta )b + \hbox{H.c.}\right]\right\}
|0\rangle_a|0\rangle_b \nonumber \;,
\end{eqnarray}
where we have used Eq.~(\ref{M}) and the transformation
\begin{eqnarray}
  S^{\dag}(\chi ){a\choose \,b^\dag } S(\chi )= \left(\matrix{\mu &\nu
\cr \nu^* &\mu \cr}\right){a\choose \,b^\dag }\;\label{M2}
\end{eqnarray}
with $\mu=\hbox{cosh}|\chi |$ and $\nu =e^{i\arg\chi }\,\hbox{sinh}|\chi |$. 
Upon defining 
\begin{eqnarray}
&&K C=e^{-i\psi_0}\mu\cos\theta
+e^{i\psi_1}\nu^*\sin\theta \;,\nonumber \\
&&KD=e^{i\psi_0}\nu^*\cos\theta +
e^{-i\psi_1}\mu\sin\theta 
\;, 
\end{eqnarray}
where $K\in \mathbb R$ and $C,D\in\mathbb C$, with $|C|^2+|D|^2=1$ one
has 
\begin{eqnarray}
K^2=\mu ^2+|\nu |^2 +2\mu|\nu |\sin 2\theta\cos(\psi_0+\psi_1-\arg
\nu)
\;. 
\end{eqnarray}
Now, since the unitary transformation 
\begin{eqnarray}
\left(\matrix{C
&D\cr D^* &C^* \cr}\right){a\choose b}\longrightarrow 
{a\choose b}\;
\end{eqnarray}
has no effect on the vacuum state, 
Eq.~(\ref{a2}) leads to the following Gaussian distribution:
\begin{eqnarray}
&&p(x;\theta,\psi_0,\psi_1)= 
\int_{-\infty}^{+\infty}\frac{\d
\lambda}{2\pi}\,e^{-i \lambda x} {}_a\langle
0|{}_b\langle 0| \exp\left\{i K\frac \lambda 2 
\left[(C\, a+ D\,b)+ \hbox{H.c.}\right]\right\}
|0\rangle_a|0\rangle_b \nonumber \\&&=
\int_{-\infty}^{+\infty}\frac{\d
\lambda}{2\pi}\,e^{-i \lambda x} {}_a\langle
0|\exp\left\{i K\frac \lambda 2 \left[a+a^\dag\right]\right\}
|0\rangle_a =\frac 1K \left|{}_a\langle 0 |x/K \rangle _{a}\right |^2 
\nonumber \\&&=
{1\over{\sqrt{2\pi\Delta^2(\theta,\psi_0,\psi_1)}}}
\exp\left(-\frac{x^2}{2\Delta^2(\theta,\psi_0,\psi_1)}\right)
\;,\label{quasifinal}
\end{eqnarray}
where the variance  $\Delta^2(\theta,\psi_0,\psi_1)$ is given by
\begin{eqnarray}
\Delta^2(\theta,\psi_0,\psi_1)=\frac {K^2}{4}=
\frac{1+|\xi|^2+2|\xi|
\sin 2\theta\cos(\psi_0+\psi_1-\arg\xi)}{4(1-|\xi|^2)}\;.
\end{eqnarray}
Taking into account the Gaussian convolution that results from
less-than-unity quantum efficiency, the variance just increases as
$\Delta^2(\theta,\psi_0,\psi_1)\to\Delta_{\eta}^2(\theta,\psi_0,\psi_1)=
\Delta^2(\theta,\psi_0,\psi_1)+\frac{1-\eta}{4\eta}$. Notice that the
probability distribution in Eq.~(\ref{quasifinal}) corresponds to a
squeezed vacuum for $\theta={\pi\over4}$ and $\psi_0+\psi_1-\arg\xi=0$
or $\pi$.
\section*{Acknowledgements}
This work has been supported by the INFM project PRA-CAT 1997 and by
the MURST Cofinanziamento program ``Amplificazione e Rivelazione di
Radiazione Quantistica''. The research of P.K. is partially supported
by the U.S. Office of Naval Research.

\newpage
\vfill\newpage
\vfill\newpage
\vfill\newpage
\vfill\newpage
\newpage
\end{document}